\documentclass[journal]{IEEEtran}
\ifCLASSINFOpdf
\else
\fi

\usepackage{datetime}
\usepackage{graphicx}
\usepackage{caption}
\usepackage{amsmath,amssymb,marvosym,wasysym}
\usepackage{adjustbox}
\usepackage{enumerate}
\usepackage{balance}
\usepackage{algorithm}
\usepackage{algpseudocode}
\usepackage{hyperref}
\usepackage{multirow}
\usepackage[table]{xcolor}
\usepackage{tabularx,colortbl}
\usepackage{balance}

\begin{document}
%
\title{Evolutionary Computing Approach to Optimize Superframe Scheduling on Industrial Wireless Sensor Networks}
%
%
%

\author{Gandeva Bayu~Satrya,~\IEEEmembership{Member,~IEEE,}
	and~Soo Young~Shin,~\IEEEmembership{Senior Member,~IEEE}\\
IT Convergence Engineering, Kumoh National Institute of Technology\\
$\{gandevabs, wdragon\}@kumoh.ac.kr$}

\markboth{Journal of \LaTeX\ Class Files,~Vol.~14, No.~8, August~2015}%
{Shell \MakeLowercase{\textit{et al.}}: Bare Demo of IEEEtran.cls for IEEE Journals}
%



\maketitle

\begin{abstract}
There has been a paradigm shift in the industrial wireless sensor domain caused by the Internet of Things (IoT). IoT is a thriving technology leading the way in short range and fixed wireless sensing. One of the issues in Industrial Wireless Sensor Network-IWSN is finding the optimal solution for minimizing the defect time in superframe scheduling. This paper proposes a method using the evolutionary algorithms approach namely particle swarm optimization (PSO), Orthogonal Learning PSO, genetic algorithms (GA) and modified GA for optimizing the scheduling of superframe. We have also evaluated a contemporary method, deadline monotonic scheduling on the ISA 100.11a. By using this standard as a case study, the presented simulations are object-oriented based, with numerous variations in the number of timeslots and wireless sensor nodes. The simulation results show that the use of GA and modified GA can provide better performance for idle and missed deadlines. A comprehensive and detailed performance evaluation is given in the paper.
\end{abstract}


%
\IEEEpeerreviewmaketitle

\section{Introduction}
\label{1}
IoT is an innovative paradigm that developed rapidly in industrial wireless systems \cite{24}\cite{25}. For instance, when IoT is applied in a smart factory scenario, an intelligent automated production system can be tracked through wireless sensor nodes, outputs can be monitored, and subsequent optimization in production can be possible for the future. Industrial wireless sensor networks consist of hundreds of sensor nodes, and the adjacent sensor nodes can have the same data \cite{1}. With the success of IWSN technology, the functionality and efficiency of wireless technology can be improved. The development of wireless technology has also developed IWSN protocol such as Zigbee, WirelessHART \cite{23}, Wireless network for Industrial Automation-Process Automation (WIA-PA) \cite{32}\cite{33} and ISA 100.11a \cite{15}. This paper will use ISA 100.11a protocol. The advantage of ISA 100.11a protocol is that it can fulfill industry needs in providing a secure and robust protocol in the communication process when applied in the field \cite{7}. However, according to previous studies, ISA 100.11a still needs some development such as optimizing time allocation for certain devices, delay restriction, and minimizing the packet drop \cite{11}.

In ISA 100.11a, the physical layer and the lower data link layer use the IEEE 802.15.4, while the upper data link layer implements TDMA (time division multiple access)\cite{1}\cite{6}\cite{7}. Based on IEEE 802.15.4, ISA 100.11a can operate either in a beacon enabled or a non-beacon enabled mode. The non-beacon enabled mode is useful for light traffic between the network nodes. The channel access and contention are performed using an unslotted CSMA-CA mechanism. In a beacon enabled network, the coordinator sends periodic beacons containing information that allows network nodes to synchronize their communications, and information on the data pending for the different network nodes. Together, all of the timeslots form what is called a superframe \cite{11}. The timeslot durations are 10 ms to 12 ms in communication networks. Thus, the schedulability of superframes becomes an interesting research topic to be investigated.

Research in recent years particularly in IWSN scheduling problems has been developed in various aspects. IETF created the 6TiSCH working group to standardize the scheduling in the IEEE 802.15.4e TSCH networks (Request for Comments-RFC 7554). In this RFC, authors discussed one of nine problems and goals in \cite{26} which scheduling mechanisms can be envisioned and could possibly coexist in IEEE 802.15.4e. Saifullah et al. \cite{27} proposed effective schedulability tests for WirelessHART networks based on a branch-and-bound technique which provides safe and reasonably tight upper bounds of the end-to-end delays of real-time flows. Duquennoy et al. \cite{28} proposed Orchestra (Robust Mesh Networks Through Autonomously Scheduled TSCH) which is to provision a set of slots for different traffic planes, and to define the slots in such a way that they can be automatically installed/removed as the RPL (routing protocol for low-power) topology evolves. Xu et al. have proposed an optional polling slots allocation method in the subnetworks to maximize the reliability and integrity of communication on WSN \cite{12}.

Furthermore, Dobslaw et al. introduced SchedEx, a generic scheduling algorithm extension which gives reliability guarantees for topologies with guaranteed lower-bounded node-to-node packet reception rates in WSN \cite{13}. Zhang et al. have proposed distributed and dynamic TDMA channel scheduling (DDCS) algorithm in WIA-PA network with multiple channels. DDCS algorithm aims to overcome interference among clusters by adopting an adaptive channel scheduling mechanism which performs channel scheduling according to actual channel conditions \cite{34}. In addition, Dinh \& Kim have evaluated the ISA 100.11a CSMA-CA by simulation \cite{14}, and Rezha \& Shin have evaluated the ISA 100.11a on three metrics, namely throughput, average delay, and energy consumption (superframe) \cite{15}. Research and analysis of ISA 100.11a is still necessary for evaluation, and analysis is needed to assist various levels of the Quality of Service (QoS) \cite{11}.

Oka and Shin \cite{6} proposed a DMS (deadline monotonic scheduling) that reduces the overhead without degrading the network performance using off-line scheduling rules and all of tasks were assumed to have static priority. Paper \cite{6} also showed that DMS can only reduce the number of beacons and can optimize miss-deadline for a limited 500ms simulation duration. In this paper, we showed that the DMS \cite{6} method in optimizing superframe scheduling still has a miss-deadline of 500ms-750ms. The problems we addressed are timeslot optimization on superframe, avoiding miss-deadline, and using evolutionary algorithms approach for RTS scenarios.

We propose that using evolutionary computing approach, which are PSO, Orthogonal Learning PSO (OLPSO), GA, and Modified GA (MGA), can optimize superframe on IWSN. PSO is part of swarm intelligence where each particle has a position vector and velocity vector. PSO \cite{29} consists of a swarm of particles, where it is initialized with a population of random solution candidates while GA aims to mimic the approach of natural evolution in computing. GA has three main components, namely selection, crossover, and mutation which is very different from PSO. It was indicated that in \cite{29} GA was very suitable for discrete cases, and since the problem in timeslot optimization on IWSN is also a discrete case, we propose the use of GA that is a metaheuristic inspired by the process of natural selection.

A modified version of GA by combining GA with DMS is presented, where DMS is used as an input to GA by using the shortest deadline for timeslot scheduling. Once the shortest deadline is retrieved through sorting, we need to enhance the timeslot by using the processes of GA namely selection, crossover, and mutation, to provide the best offspring (solution). In practical, GA directly evaluates each candidate solution using its defect time and calculates the corresponding fitness value. Then, by using object-oriented simulations, we compare EDF (earliest deadline first) \cite{30}, DMS \cite{6}, PSO, OLPSO, GA, and MGA with variations on time duration and sensor nodes. We list our contributions as follows:
\begin{enumerate} 
	\item We demonstrate the drawback of DMS algorithm, which follows the network model and system model of ISA 100.11a.
	\item We propose an optimal GA for optimizing superframe scheduling in IWSN with detailed configuration.
	\item We define algorithms that are the transformation of TDMA based on the genetics form (genotype and phenotype) and simulated it based on object-oriented programming.
	\item We evaluate and prove that GA and MGA outperform the existing benchmark techniques (EDF, DMS, PSO, and OLPSO).
\end{enumerate}

The rest of this paper is organized as follows. Section \ref{2} will review relevant literatures. Section \ref{3} discusses the design of the proposed system that uses GA based on ISA 100.11a standard. Section \ref{4} presents the case study and simulation results. Section \ref{5} consists of the analysis and discussions. Section \ref{6} suggests future work and extensions for this paper. Finally, section \ref{7} provides the conclusion and future work from this paper.

\section{Literature review}
\label{2}
\subsection{Scheduling in ISA 100.11a}
Dewanta et al. \cite{9}, proposed message scheduling on the dedicated time slot of ISA 100.11a to satisfy real time property. A superframe is designed to accommodate periodic real time (PRT) messages, aperiodic real time (alarm) messages, and non-real time (NRT) messages. Superframe will be divided into dedicated time slot (DTS) which is used to send PRT messages and shared time slot (STS) which is used to send alarm and NRT messages. This approach is achieved by employing superframe and multi-superframe to accommodate those PRT-messages without neglecting NRT and alarm messages. However, the results show that using the message scheduling technique requires more beacons in the system and increases overhead.

Oka and Shin \cite{6} proposed a DMS's scheme that has been used to check schedulability of superframe with beacon-enabled mode in ISA 100.11a networks. A new application of deadline monotonic scheduling is proposed to check and test superframe scheduling and to reduce the overhead without degrading the network performance in ISA 100.11a IWSN. The simulation results showed that the proposed method required fewer beacons, compared to message scheduling. However, that scenario only works in the case of four sensors and static parameters (release, computation, deadline, and periodic time).

A different approach from Nhon and Kim proposed two new message scheduling methods on shared timeslots of the ISA 100.11a standard to enhance real-time performance, namely, traffic-aware message scheduling (TAMS) and contention window size adjustment (CWSA) \cite{8}. In TAMS, instead of competing to transmit sporadic messages in consecutive cycles, end-nodes are divided into groups, which then access the channel in specific cycles when the probability of timeslots getting involved in collisions exceeds a specified threshold. Conversely, in CWSA, the contention window is adjusted when the probability of timeslots getting involved in collisions exceeds the threshold. The results of simulations conducted indicate that these two proposed methods provide performance improvements in terms of success probability and end-to-end delay. However, this paper can only be applied to star topologies and limited simulation time.

\subsection{Artificial intelligence based}
Meanwhile using the artificial intelligence approach, Norouzi et al. investigated GA as a dynamic technique to find optimum states for lifetime and energy consumption of WSNs \cite{4}. These two competing objectives have a deep influence over the service qualification of networks and according to recent studies, cluster formation is an appropriate solution for their achievement. To transmit aggregated data to the Base Station (BS), logical nodes called Cluster Heads (CHs) are required to relay data from the fixed-range sensing nodes located on the ground to high altitude aircraft. The simulation diagrams indicate that using a GA-based cluster formation algorithm extends the lifetime of the network through equally distributed clustering. However, in this paper there is no packet priority scenario and the environment used was WSN.

Ziari et al. \cite{5}, proposed a hybrid of GA and PSO based on TDMA Scheduling in wireless sensor networks. Protocol TDMA in these networks is a suitable one for saving nodes energy. In this protocol, time was divided into time slots of equal length. The purpose of TDMA scheduling is to assign time slots into nodes in a way that minimizes the total number of time slots for sending data packages and energy consumption. However, in this paper there is no packet priority scenario and the environment used was WSN.

Other papers state that GA has been successfully applied in WPS (Weapon Production Scheduling)  in the category of flexible job shop scheduling \cite{2}. Furthermore, GA has also helped in container handling operations at the Patrick AutoStrad container terminal located in Brisbane Australia \cite{3}. Along with those four papers that used GA as their basis \cite{2}\cite{3}\cite{4}\cite{5}, this paper would also apply the ISA 100.11a using GA in the optimization of packet scheduling of multiple sensors with different priorities and deadlines as the most important attribute.

\begin{table}
	\caption{Simulation Parameters for IWSN Scheduling}
	\label{tab:1}
	\centering
	\begin{tabular}{| l | c | c | c | c |}
		\hline
		\textbf{Nodes} & \textbf{$r_{M_{n}}$ (ms)} & \textbf{$c_{M_{n}}$ (ms)} & \textbf{$d_{M_{n}}$ (ms)} & \textbf{$t_{M_{n}}$ (ms)} \\ \hline\hline
		Beacon & 0 & 10 & 10 & 250\\ \hline
		Node 1 & 10 & 20 & 20 & 150\\ \hline
		Node 2 & 20 & 20 & 80 & 80\\ \hline
		Node 3 & 30 & 30 & 100 & 100\\ \hline
		Node 4 & 40 & 10 & 50 & 50\\ \hline
	\end{tabular}
\end{table}

\begin{table*}
	\caption{Simulation upto three superframes; Error in Oka \& Shin}
	\label{tab:2}
	\footnotesize
	\centering
	\begin{tabular}{ l l c c c c c c c c c c c c c c c }
		\hline
		\textbf{Slot number} 	& [1] 			& \textbf{...} 	& [43]	& [44] & [45] & [46] & [47] & [48] & [49] & [50] & [51] & [52] & [53] & [54] & \textbf{...} & [75]\\ \hline\hline
		\textbf{Executed node} 	& 0 			& ... 	  	& 2		& 2 & 4 & 3 & 1 & 1 & 3 & 4 & 0 & 2 & 2 & 3 & & 4 \\ \hline\hline
		Beacon 0 				& 0 			& ... 	  	&		&  &  &  &  &  &  &  & 0 &  &  &  &  &\\ 
		Node 1					&  				& ... 		&		&  &  &  & 1 & 1 &  &  &  &  &  &  &  &\\ 
		Node 2 					&  				& ... 		& 2		& 2 &  &  &  &  &  &  &  & 2 & 2 &  &  &\\ 
		\rowcolor{cyan}
		Node 3 					&  				& ... 		&		&  &  & 3 &  &  & 3 &  &  & \DOWNarrow & \DOWNarrow & 3 &  &\\ 
		Node 4 					&  				& ... 		&		&  & 4 &  &  &  &  & 4 &  &  &  &  & & 4 \\ \hline\hline
		\textbf{Missed Deadline} 	& \multicolumn{4}{l}{1 (80 ms) for node 3} 			  &  &  &  &  &  &  &  &  &  &  &  &\\ 
		\textbf{Idle} 			& \multicolumn{4}{l}{8 (80 ms)} 			  &  &  &  &  &  &  &  &  &  &  &  &\\ 
		\textbf{Total Defect time} & \multicolumn{4}{l}{160 ms} 			  &  &  &  &  &  &  &  &  &  &  &  &\\  \hline \\
		\multicolumn{6}{l}{\textbf{Note}: \textit{Symbol \DOWNarrow, means there is a missed deadline}.}
	\end{tabular}
\end{table*}

\section{Design topology and algorithm}
\label{3}

\subsection{Network model}

\begin{figure}[h]
	\centering
	\includegraphics [width=0.35\textwidth]{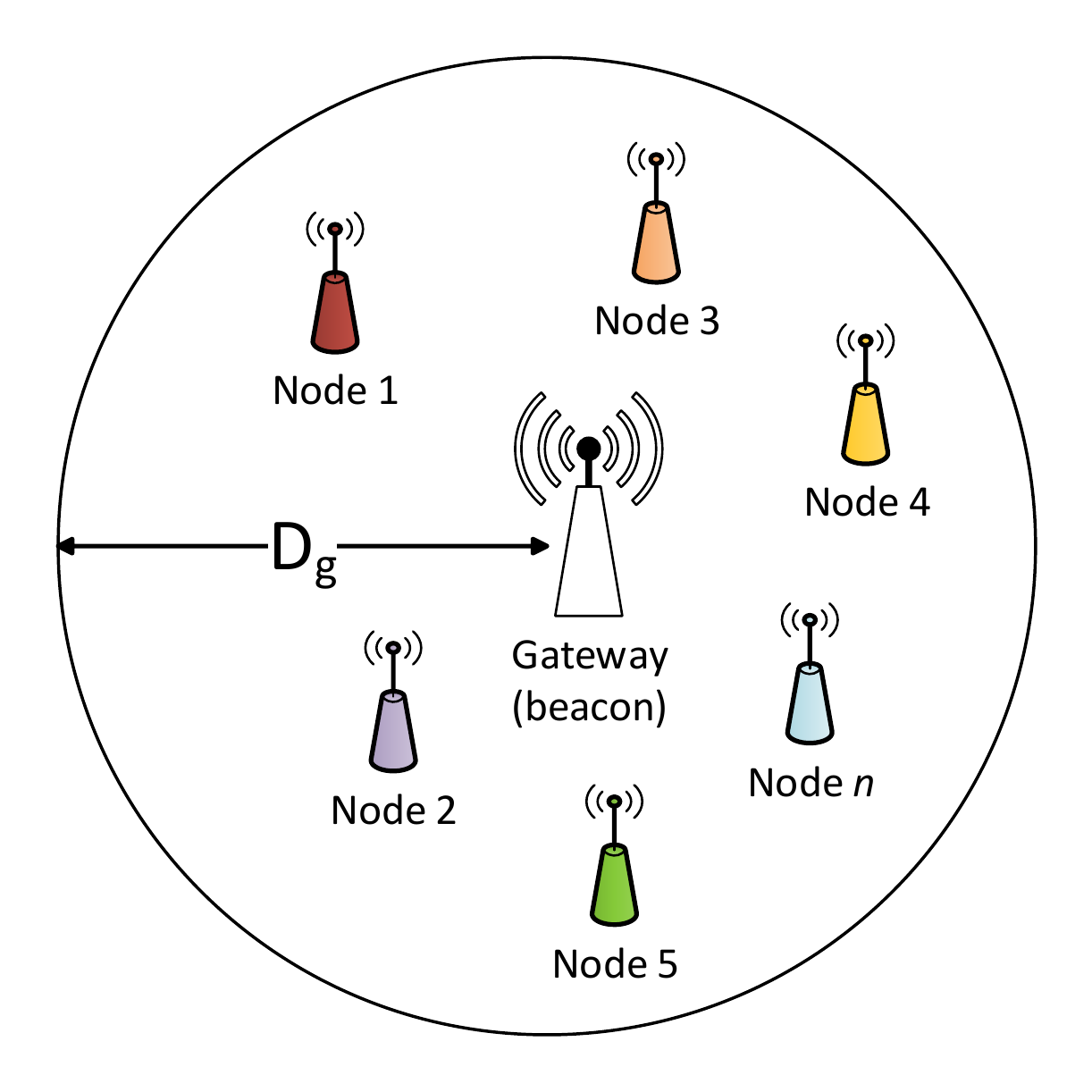} \\
	\caption {network model for IWSN}
	\label{fig:1}
\end{figure}

Our proposed scheme is an improvement and development of \cite{6}. For network modeling adopted from \cite{6}, there has been little development of the wireless sensor nodes and the amount used is up to ten nodes. This paper focuses on single-channel and adopts star networking topology, as seen in Fig. \ref{fig:1} which is the maximum area of a gateway is $D_g$ and sensor nodes vary from node-1 until node-n. Each wireless sensor node represents different sensors and information as shown in Table \ref{tab:1}. The main problem that we want to show is that each wireless sensor node has periodic transmission and deadline generated by the node due to the unpredictable bursty nature of the traffic. The use of genetic algorithms to solve the problems of scheduling is based on the deadline of each sensor. A service network model is created to serve a monitoring system in industrial areas. As described in \cite{6}, we also adopt a standard for industrial wireless namely ISA 100.11a as a sample case study.

Table \ref{tab:1} is the detailed information of each node on the network topology. The scenario starts data traffic on the gateway which acts as a coordinator that will send a beacon periodically on the network. We assume that $Mn$ is a periodic message for each node, $Mn = (M_{nB}, M_{n1}, M_{n2}, M_{n3}, ..., M_{n10})$. $M_{nB}$ is a periodic message for gateway that send beacons.  $M_{n1}$ is a periodic message for node 1. In this case study, the beacon is the highest priority in the calculation of deadlines. In short, we place a beacon as the top priority in the calculation of the conditional expression.

\subsection{Review of DMS method in \cite{6}}
\label{32}

\begin{figure}[h]
	\centering
	\includegraphics [width=0.5\textwidth]{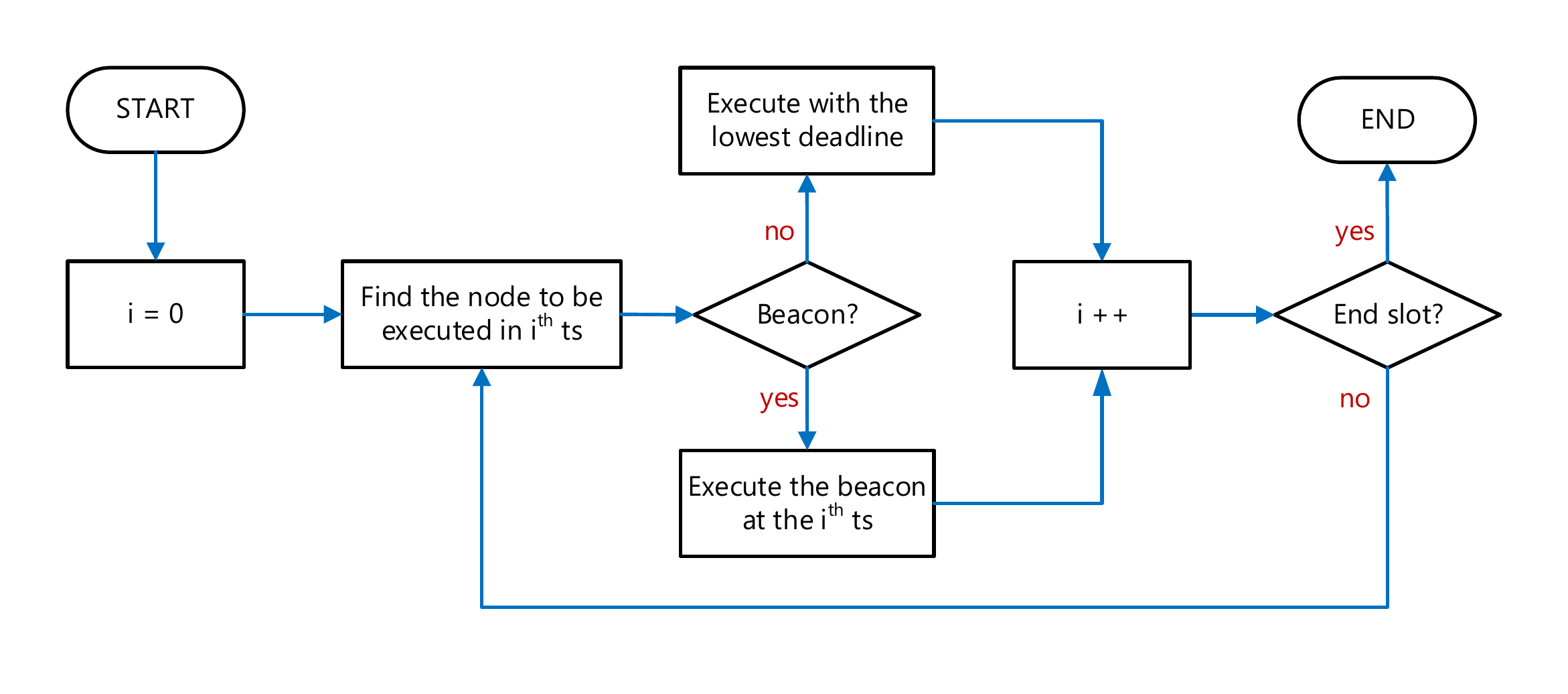} \\
	\caption {DMS flowchart}
	\label{fig:2}
\end{figure}

In Oka \& Shin's paper, DMS (as can be seen on Fig. \ref{fig:2}) is used for reducing the number of beacons, which is an improvement on the method used by the message scheduling by Dewanta et al. The scenarios used in the paper \cite{6} were only valid for one superframe, where one superframe consists of 25 timeslots (ts) and 1 ts = 10ms. Table \ref{tab:2} shows three superframes, $M_{n3}$ (message for the nodes 3) is delayed between the second and third superframe. In accordance with the data in the Table \ref{tab:1}, node 3 has 30 ms of computation time and 100 ms of deadline time. Hence, node 3 should be executed at ts-52 or ts-53. Ts-54 is the fifth period \textit{(the allocation of subsequent periods)} for node 3. As seen in Table \ref{tab:2} with DMS, node 3 was only executed twice, while it should be executed 3 times in one period. In summary, node 3 was delayed in the second superframe (ts-52 and ts-53).

Based on this outcome (as presented in Table \ref{tab:2}), DMS is not schedulable for the superframe scheduling test. Hence, DMS is not suitable when applied to a real-time case. In addition, on IWSN, especially ISA 100.11a such systems will be run for 24 hours every day. The proposed solution of this paper uses genetic algorithm combined with DMS to build a more intelligent IWSN. It will compare and evaluate EDF, DMS, traditional GA, traditional PSO and the latest version of PSO which is OLPSO.

\subsection{Proposed algorithm}
\label{pro}

\begin{figure}[h]
	\centering
	\includegraphics [width=0.45\textwidth]{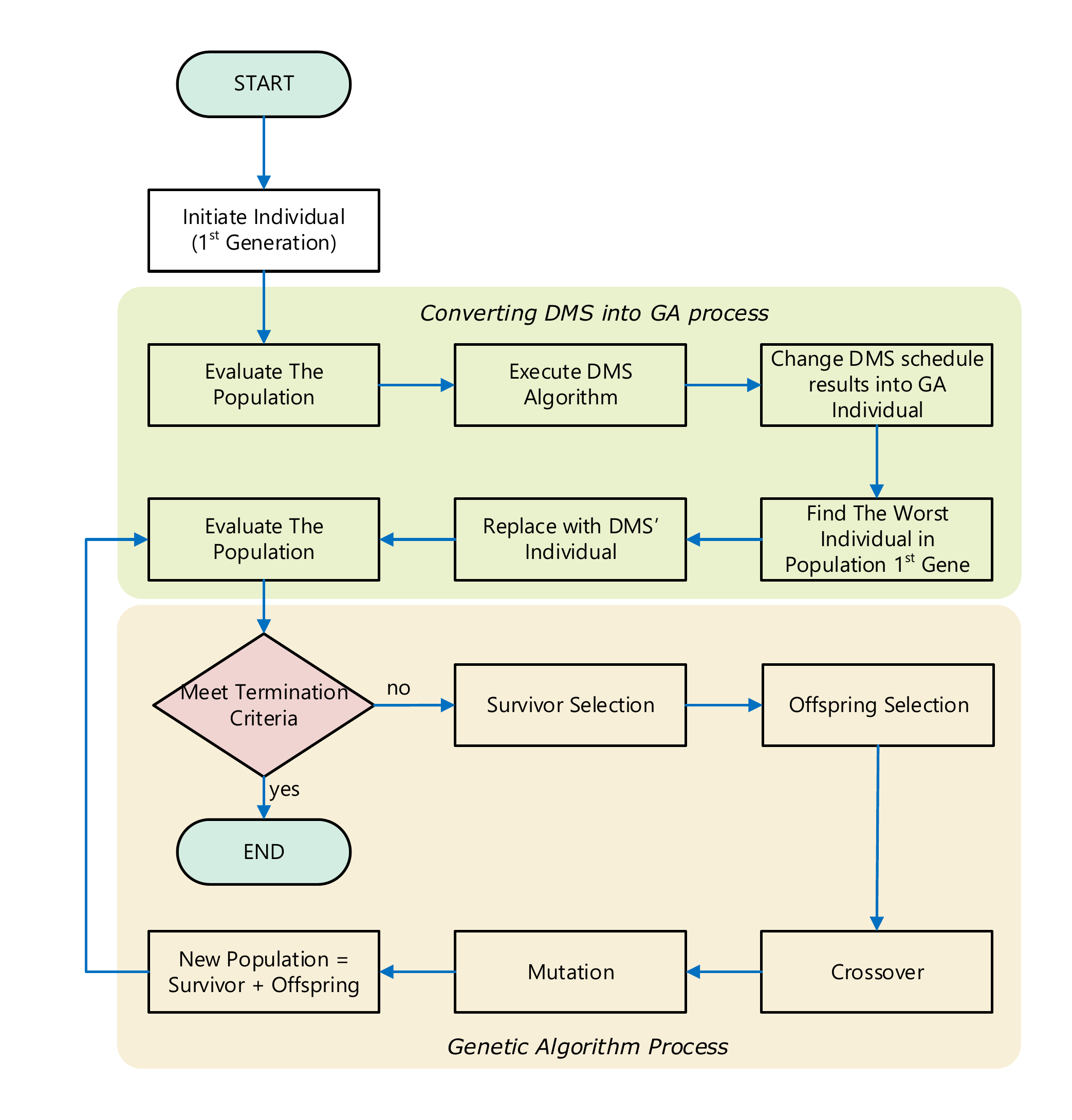} \\
	\caption {MGA flowchart}
	\label{fig:3}
\end{figure}

According to \cite{22}'s recommendations in the book entitled "Adaptation in Natural and Artificial Systems", any problems in the form of adaptation can be formulated in biological processes. Based on this, we are trying to modify GA by adding DMS algorithm as an aid to solve problems in the IWSN superframe scheduling. Our proposed MGA is presented in Algorithm \ref{isa} and the design flowchart in Fig. \ref{fig:3}. We put the DMS at the beginning of population establishment. The best results of DMS which are sorted by the smallest deadline will be the input of GA \textit{(as can be seen in Algorithm \ref{isa} step 6)}. Reasons for modifying GA is that the best population will be acquired from the best parents in the biological process. It is expected to assist in optimizing the timeslots and missed deadlines in a superframe scheduling case \textit{(as illustrated in Algorithm \ref{isa} step 14)}. The notations used in Algorithm \ref{isa} are as follows. $P_g$ is the $g^{th}$ population index, $I$ is an individual that has been obtained from DMS, $S_g$ is a collection of individual survivors in the index $g$, and $O_g$ is a collection of offspring individuals in the index $g$.

\begin{algorithm}
	\caption{MGA for ISA Scheduling}
	\label{isa}
	\begin{algorithmic}[1]
		\State INITIALISE g $\leftarrow$ 1;								\Comment{initiation of generation counter}
		\State INITIALISE $P_g \leftarrow P_{initial}$;					\Comment{generate individuals for $1^{st}$ population}
		\State EVALUATE ($P_g$);										\Comment{calculate the fitness value}
		\State CONVERT $I \leftarrow ConvertScheduler (DMS_{schedule})$;\Comment{transform the DMS becoming phenotype \& genotype}
		\State REPLACE $P_g \leftarrow ReplaceWorstIndividual (P_g, I)$;\Comment{substitute with new individual (I)}
		\State EVALUATE ($P_g$);										\Comment{calculate the fitness value}
		\While {!finished}												\Comment{meet termination criteria}
		\State $g \leftarrow g+1 $										\Comment{increment generation counter}
		\State $S_g \leftarrow Select (P_{g-1}, N)$						\Comment{Selection process for survivor}
		\State $O_g \leftarrow Select (P_{g-1}, P_{g-1}.Count - N)$		\Comment{Selection process for offspring}
		\State $O_g \leftarrow Crossover (O_g)$							\Comment{Recombination process, using single point crossover}
		\State $O_g \leftarrow Mutation (O_g)$							\Comment{Mutation process, using random mutator}
		\State $P_g \leftarrow S_g+O_g $								\Comment{New population is created based on survivor and offspring}
		\State EVALUATE ($P_g$);					
		\EndWhile
	\end{algorithmic}
\end{algorithm}

\subsubsection{Genetic representation}

The first stage in building a GA is to decide the genetic representation for the population along with the candidate solution to a problem. It involves the design of genotype and phenotype. The first step from the point of view of automated problem-solving is to decide how possible solutions should be specified and stored in a way that can be manipulated by a computer. We say that objects forming possible solutions within the original problem context are referred to as phenotypes, while their encoding, that is, the individuals within the EA (Evolutionary Algorithms), are called genotypes.

In many cases, there will be plenty of choices and to get proper representation is the hardest thing in designing GA. In this case study, we use integer representation as shown in Fig. \ref{fig:6} and we also define the values in Table \ref{tab:1}. Individuals are represented in integer. The integer in each gene expresses the index number of the nodes, in addition to the beacon, the length of genes \textit{(number of timeslots - timeslots for beacon)}, and the position of the gene sequence signifies the execution.

\begin{figure}[h]
	\centering
	\includegraphics [width=0.38\textwidth]{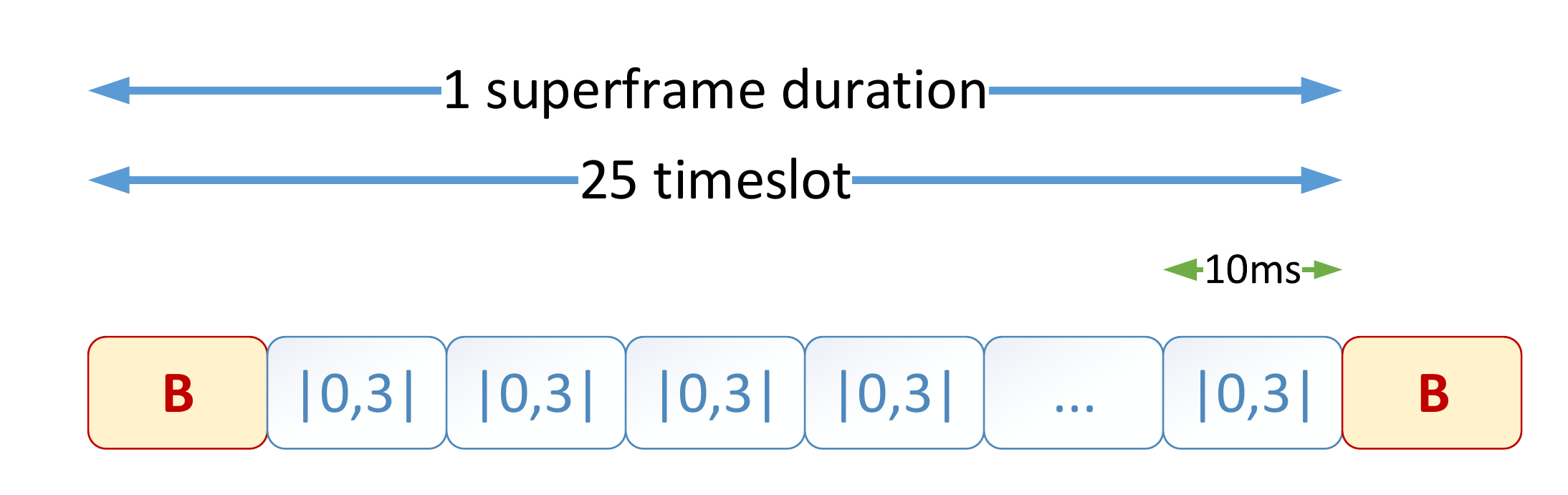} \\
	\caption {Representation of superframe}
	\label{fig:4}
\end{figure}

Based on previous exposure, the gene length = 4 for slots for the beacon (B) = 0, i.e. at the first slot, and the possible values that can be filled on each gene between the grades 0-3 are declared node 0 for index 0, the nodes one for index 1, and so on. So, that individual representation is as shown in Fig. \ref{fig:4}.

\subsubsection{Fitness function}
Fitness function is the function given to calculate the quality of one (or multiple) genotypes. The role of this function is to represent the fitness of the population requirements that should describe the optimal convergence rate. The population must adapt to meet the criteria indicated by the evaluation function. Simplifying the real world evolution concepts, individuals with high fitness value will survive, while those with a low fitness value will die. Each feasible solution can be characterized by the fitness value for the problems.

In this case, we want the best scheduling. We define that the best schedule has the smallest defect time ($d_t$). Defect time is the sum of idle time ($id_t$) and lateness time ($l_t$). Genetic representation does not apply to the beacon, because according to the requirements of data transfer on the IEEE 802.15.4 the beacon is a particular priority in the synchronization phase \cite{11}. Based on Eq. \ref{eq:1} if the defect time = 0, where there are no idle and missed deadlines, then the value of fitness in Eq. \ref{eq:2} is 1. The expected value of fitness is no delay or idle in each timeslot.

\begin{equation}
\begin{aligned}
\label{eq:1}
d_t = id_t + l_t
\end{aligned}
\end{equation}

So, the fitness value is

\begin{equation}
\begin{aligned}
\label{eq:2}
f(d_t) =
\begin{cases}
1                & ,d_t = 0 \\
\frac{1}{d_t}    & ,d_t > 0
\end{cases}
\end{aligned}
\end{equation}

\begin{figure}
	\centering
	\includegraphics [width=0.5\textwidth]{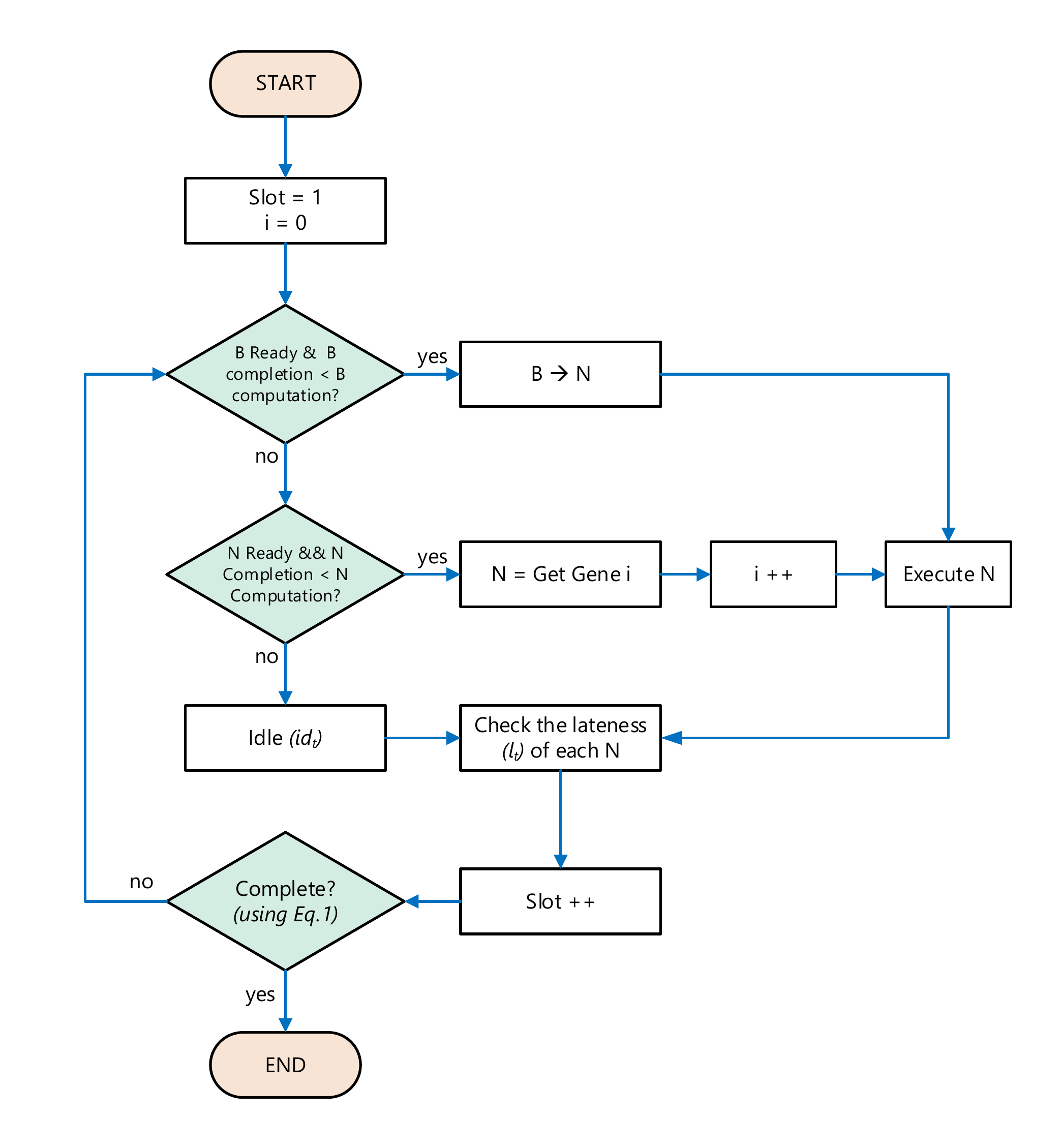} \\
	\caption {Flowchart for calculating defect time}
	\label{fig:5}
\end{figure}

Fig. \ref{fig:5} is the implementation of \textit{'defect time'} calculation on GA as a fitness value. Based on Eq. \ref{eq:1}, which calculates the missed deadline values, different parameters are considered such as  $r_{M_n}$ for release time, $d_{M_n}$ for deadline time, $t_{M_n}$ for periodic time, and $c_{M_n}$ for computation time (as shown in Table \ref{tab:1}). Now to calculate these missed deadlines, this paper follows the theorem proposed in \cite{6} which states that superframe can only be scheduled if comply to $ \forall{M_n} : c_{M_n} \leq d_{M_n}$.

The \textit{N} and \textit{B} notations are nodes to be executed and beacon, respectively. Completion time is the total of time needed for the system for executing nodes. Based on \cite{9} beacon timeslot is used by gateway to inform timeslot occupation to all nodes which has been illustrated in Fig \ref{fig:4}. More explanation for calculating the defect time can be seen in Fig \ref{fig:5}. First, initially we set i = 0 and slot = 1. Second, if the beacon is ready and has not been finished executing, then execute beacon on the current timeslot. Third, If node at the $i^{th}-gene$ is ready and has not been finished executing, then the node will be executed on the current timeslot and increment \textit{i}. Fourth, otherwise, set idle on the current timeslot. Fifth, Check if there is tardiness on the node. Sixth, increment the timeslot. Seventh, If all genes have been executed, then end the process. Otherwise, go to second step again.

\subsubsection{Encoding genotype}

\begin{figure}[h]
	\centering
	\includegraphics [width=0.45\textwidth]{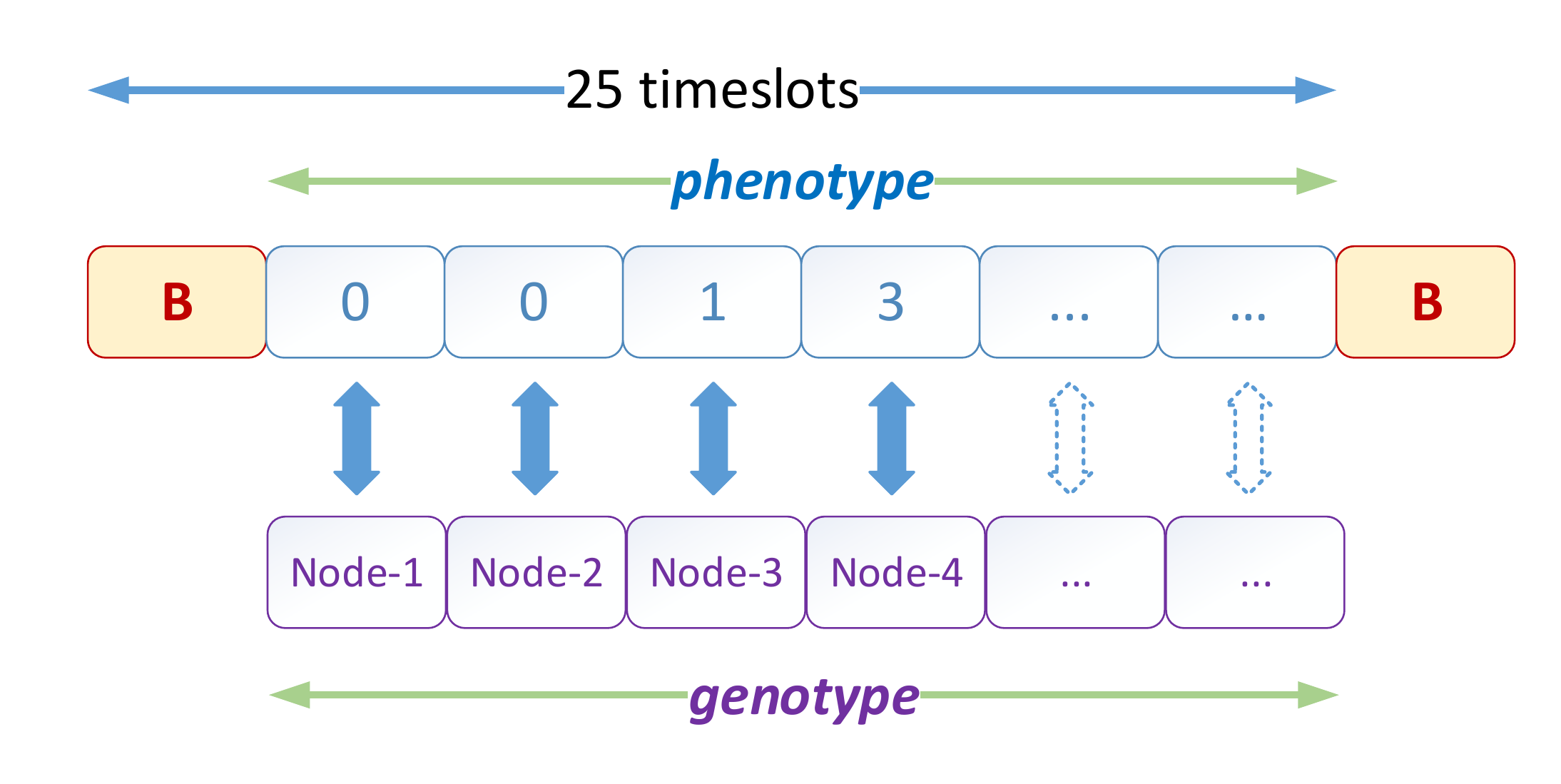} \\
	\caption {Representation of phenotype \& genotype}
	\label{fig:6}
\end{figure}

Based on the case presented in individual representation, if there are individuals, for example $\mid0\mid 0\mid 1\mid 3\mid$, then the resulting schedule is as shown in Fig. \ref{fig:6}.

\subsubsection{Recombination}
Recombination is a process of establishing a new individual for the information contained in the two (or more) parent solutions. Recombination is often also called a crossover, which is motivated by the analogy of biology. Recombination operators are applied based on the crossover rate $p_c$. Suppose two elected parents and two offspring are produced in the recombination probability of two parents with a $p_c$. In other words, copying parents with probability $1-p_c$. Integer representation could be using one-point crossover, n-point crossover and uniform crossover \cite{10}. All of these types, will be tested on this system. Not all of the chromosomes in a population will sustain the process of recombination. The possibility of a chromosome sustaining the recombination process is based on the probability of crossover that has been predetermined.

\subsubsection{Mutation}
The mutation process occurs after the process of recombination by selecting chromosomes that will be mutated randomly, and then determining the point of mutation in the chromosomes randomly. The number of mutated chromosomes will be calculated based on the probability of mutation $p_m$ that has been predetermined. Selecting the position of the mutated genes, can be done by generating a random integer number, between 1 to \textit{total\_gen}. If the random number generated is smaller than the variable \textit{mutation\_rate} $p_m$ then it selects the position as a sub-chromosome to be mutated. For example if $p_m$ is set at 10\%, it is expected that 10\% of a \textit{total\_gen} will be mutated.

Integer representation, can use the 'random resetting' and 'creep mutation' which will mutate each gene independently with probability $p_m$ predetermined by the user. These systems will use the 'random resetting' because 'creep mutation' is only used for ordinal attributes whereas this case is in the form of cardinal attributes. For further explanation about ordinal and cardinal attributes refer to \cite{10}.

\subsubsection{Termination condition}
GA and other EA are stochastic and mostly no warranty will achieve optimal value, there could be a situation that this condition will never be reached, and the algorithm can not be stopped \cite{10}. Therefore we have to design and predict if these conditions will occur and how to create a condition in which the algorithm will stop. In our system, searching for a new individual process will stop if the fitness value does not increase again on 1000 generations.

\section{Case study and simulation}
\label{4}

We have already explained DMS in sub-section \ref{32}. In this section, we focus on GA and MGA. We also compare all algorithms with EDF as other representative of non-metaheuristic. Based on the details of the parameters in Table \ref{tab:1} and the topology in the Fig. \ref{fig:1}, the following is a configuration of a GA that we used. Explanation of the terms we use in Table \ref{tab:3} are given in sub-section \ref{pro} above. As a comparison in EA classification we use PSO with 50 particles  as an evaluation. During the iteration time t, the update of the velocity from the previous velocity to the new velocity is determined by Eq \ref{eq:3}. The new position is then determined by the sum of the previous position and the new velocity by Eq. \ref{eq:4}.

\begin{equation}
\label{eq:3}
\begin{split}
v_i(t) = (1-\frac{t}{m}) v_i(t-1) + c_1r_1(Pos_{pBest}-Pos_{t-1}) +
\\
c_2r_2(Pos_{gBest}-Pos_{t-1})
\end{split}
\end{equation}

\begin{equation}
\label{eq:4}
Pos_t= v_t + Pos_{t-1}
\end{equation}

where,

\begin{tabular}{ l l } 
\textit{v} 		&: \textit{velocity }						\\ 
\textit{t}		&: \textit{iteration}					\\
\textit{m}		&: \textit{maximum iteration (1000) }		\\
\textit{c}		&: \textit{constant ($c_1=2, c_2=2$) }		\\
\textit{r}		&: \textit{random value (0-1) }				\\
\textit{Pos}	&: \textit{position} 						\\
\textit{pBest}	&: \textit{best particle per iteration} 		\\
\textit{gBest}	&: \textit{best particle on the overall iteration} 	\\ \\
\end{tabular}

For a more comprehensive comparison, OLPSO will be used as the latest development of the PSO \cite{43}\cite{44} family. In OLPSO, the new mechanism is introduced as the OED \textit{(orthogonal experimental design)} process. The processes at OED replace the learning processes in traditional PSO with calculation as follows:

\begin{equation}
\label{eq:5}
v_id = (1-\frac{d}{m}) v_i(d-1) + cr_d(p_{od}-x_{id})
\end{equation}

where, the current particle $X_i$ is between its personal best position $P_i$ and its neighborhood’s best position $P_n$. Vector $P_o$ stores only the index of $P_i$ and $P_n$, not the copy of the real position values. That is, $p_{od}$ only indicates that the $d^{th}$ dimension is guided by $P_i$ or $P_n$. The OLPSO simulations still used 50 particles and 1000 iterations.

These simulations were run using Java \cite{21} with JDK 1.8.0\_151 as the implementation of its development. For evolutionary algorithms and the genetic programming library, we used Jenetics (jenetics.io). Whereas for the programming editor, we have used NetBeans 8.2 64bits for Windows. The simulations that have been built used a variation of timeslot numbers i.e. 100, 200, and 500. For nodes count, we used 1 node as the beacon while the number of wireless sensor nodes vary between 4, 7, 10, and 100 nodes.

Results from the test with four sensor nodes and one beacon were explained in Table \ref{tab:2} above on DMS review. We proved that DMS failed to handle miss deadline on the third superframe. Here were some sample results for 7, 10, and 100 sensor nodes. The total results of this experiment would be presented in section \ref{5}.

\begin{table}[]
	\centering
	\caption{Genetic algorithm configuration}
	\label{tab:3}
	\begin{tabular}{ | l | l | } \hline
		\textbf{Parameter}  	& \textbf{Value} \\ \hline \hline
		GA population size 		& 50-100 \\ \hline
		Number of generations  	& 1000 iterations \\ \hline
		Selection method  		& Tournament / Truncate  \\ \hline
		Offspring Fraction 		& 0.4 \\ \hline
		Crossover method		& Single Point Crossover (0.2) \\ \hline
		& Multi Point Crossover (0.2 / 2 - 10) \\ \hline
		Mutation method  		& Gaussian Mutator (0.00001 - 0.001) \\ \hline
		Max Phenotype Age  		& No Phenotype Age \\ \hline
	\end{tabular}
\end{table}

\subsection{Simulation result for 7 nodes}

\begin{figure}
	\centering
	\includegraphics [width=0.48\textwidth]{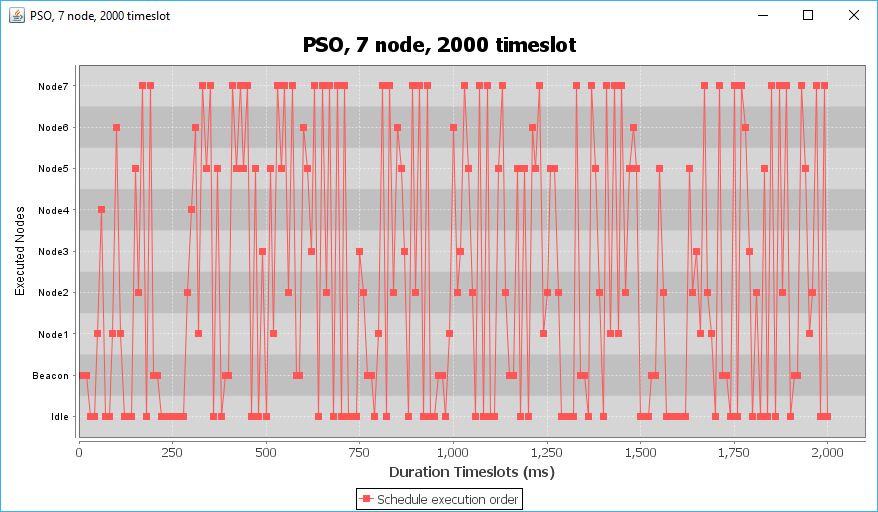} \\
	\caption{Simulation: PSO with 200 timeslot and 7 nodes}
	\label{fig:7}
\end{figure}
\begin{figure}
	\centering
	\includegraphics [width=0.48\textwidth]{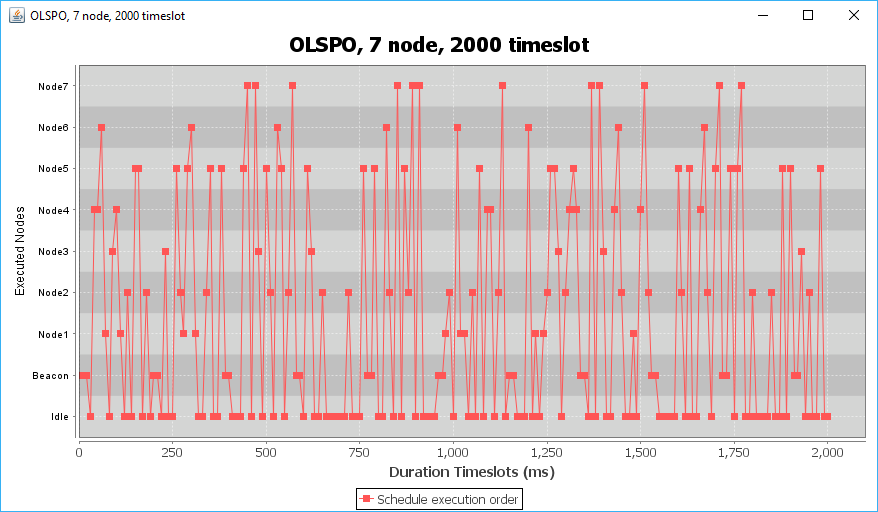} \\
	\caption{Simulation: OLPSO with 200 timeslot and 7 nodes}
	\label{fig:8}
\end{figure}
\begin{figure}
	\centering
	\includegraphics [width=0.48\textwidth]{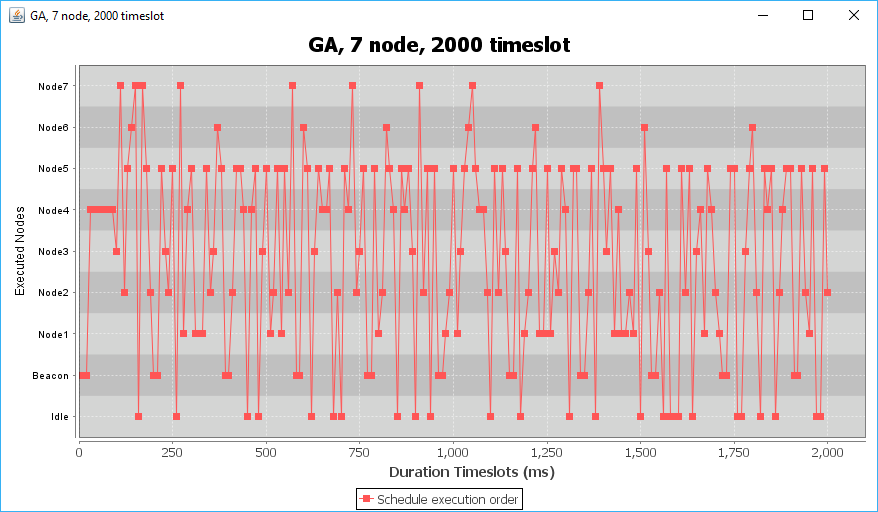} \\
	\caption{Simulation: GA with 200 timeslot and 7 nodes}
	\label{fig:9}
\end{figure}
\begin{figure}
	\centering
	\includegraphics [width=0.48\textwidth]{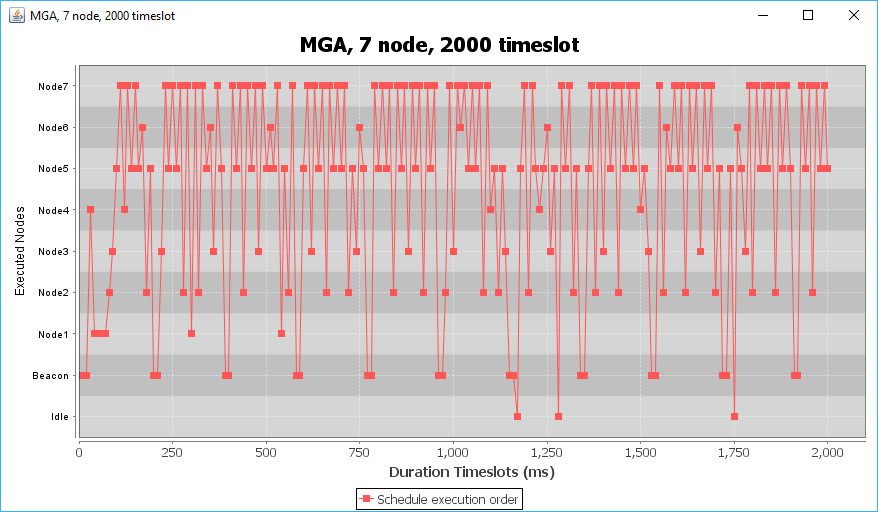} \\
	\caption{Simulation: MGA with 200 timeslot and 7 nodes}
	\label{fig:10}
\end{figure}

Below are four results (Figs. \ref{fig:7}, \ref{fig:8}, \ref{fig:9}, and \ref{fig:10}) of 60 scenarios that were carried out. This experiment used one beacon, seven wireless sensor nodes, 200 ts (or 2000ms). Fig. \ref{fig:10} is a simulation of results using MGA. When compared with traditional PSO, OLPSO and GA, MGA produces much better defect-time performances.

Basically, PSO has a different character from GA. In PSO, particle refers to one solution which according to it is better, then it will change its direction. On the contrary, if there is no solution found, then it will keep the previous vector. This timeslot optimization case is a combinatorial and discrete case. If we look at the characteristics of PSO which is continuous, PSO will convert its iteration results into discrete form. However, the characteristics of GA are very suitable for combinatorial cases, if a bad gene were found, then it can do selection, mutation, or crossover.

The characteristics of OLPSO are slightly different with traditional PSO in term of searching the best position with OED \textit{(orthogonal experimental design)}. As seen in Fig. \ref{fig:8}, the defect time of OLPSO is better than that of PSO but worse than the defect time of EDF, DMS, GA and MGA. This is because of too many particles and generations in narrow searching space that is 7 sensor nodes. In other words, OLPSO is not suitable for simple cases with low number of sensors and small timeslots.

In addition, we make modifications to the GA by combining DMS with GA or what we call MGA. Based on the simulation, we found that the MGA is much better than conventional DMS or conventional GA. This is because in the evaluation for population, we combined the output of DMS as a new individual for GA. This means that GA has a population with a good fitness, and so when those good individuals are mated (crossover) with other individuals, the results would be better than before the mating.

\subsection{Simulation result for 10 nodes}

\begin{figure}
	\centering
	\includegraphics [width=0.48\textwidth]{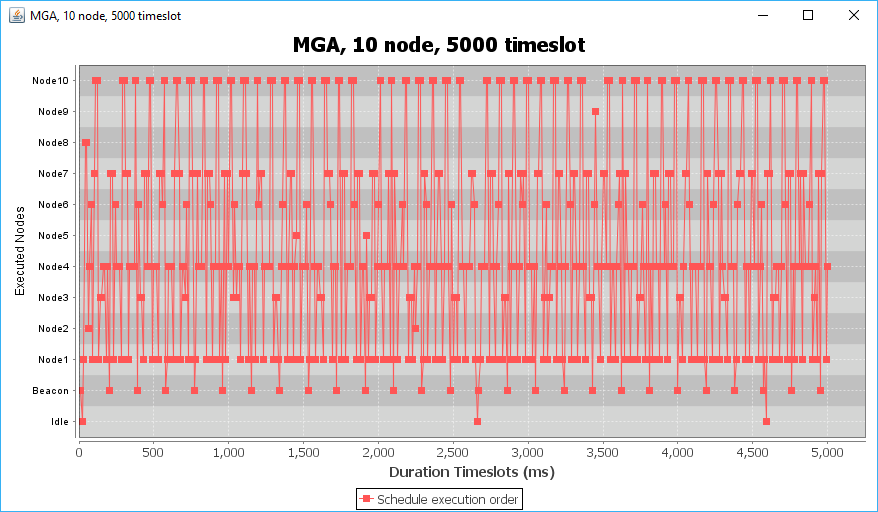} \\
	\caption{Simulation: MGA with 500 timeslot and 10 nodes}
	\label{fig:14}
\end{figure}

The following Figs. \ref{fig:11}, \ref{fig:12}, \ref{fig:13}, and \ref{fig:14} are the four scenarios' test results out of 60 scenarios' tests that were carried out. This experiment used one beacon, 10 wireless sensor nodes, 500 ts (or 5000 ms). With the same description as in the previous explanation, in this test, the number of nodes given were more complex. The purpose of this test is to determine how well the GA is able to perform optimization on the superframe scheduling. As previously explained, the parameter for sensor nodes such as release time, computation time, deadline time, and periodic time will be generated randomly. In short, GA and MGA are capable of handling sensor data in any quantity, with the condition to keep watching the fitness values and parameters of the GA that will be set (see Table \ref{tab:3}). This way, it can help to discover a better solution for GA that is categorized with stochastic algorithms. Stochastic is a random occurrence where the appearance of an individual cannot be predicted \cite{10}, however, when measured from the distribution throughout the observation, it will usually follow a pattern. From here, it requires accuracy in the individual selection, mutation and crossover.

Fig. \ref{fig:11} shows that PSO was not optimal in optimizing the timeslot because there were still many empty 'ts'. If we look at idle time, PSO has 266 ts, GA has 105 ts, and MGA has 3 ts. Idle time is one of the evaluation metric of fitness value in the most optimal calculation. In these cases, the timeslot showed that PSO may easily get trapped in local optima which led to bad solution. Besides, MGA has been proved to be the best for combinatorial cases like this compared to PSO and GA.

However, the experiment using 10 SN showed that OLPSO is worse than EDF, DMS, PSO, GA, and MGA. Without the loss of generality, the mechanism on OLPSO, based on Equation \ref{eq:5}, uses floating point arithmetic. As explained in X, $p_{od}$ is a guidance vector consists of combinations of $P_i$ and $P_n$. In every generation, particle i will update its own velocity and position. To the best of authors knowledge, in the OLPSO algorithm, when $P_i$ or $P_n$ changes to a better position, the new information will be adopted immediately by the particle through $P_o$.

\begin{figure}
	\centering
	\includegraphics [width=0.48\textwidth]{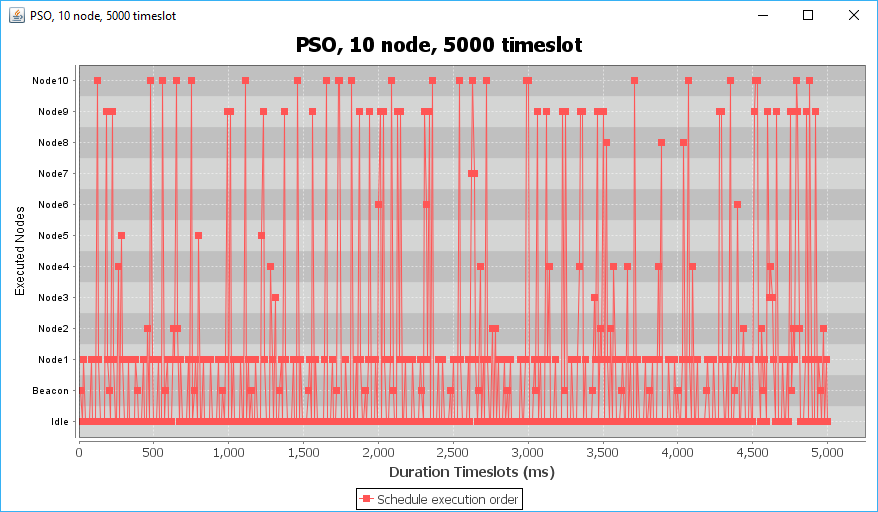} \\
	\caption{Simulation: PSO with 500 timeslot and 10 nodes}
	\label{fig:11}
\end{figure}

\begin{figure}
	\centering
	\includegraphics [width=0.48\textwidth]{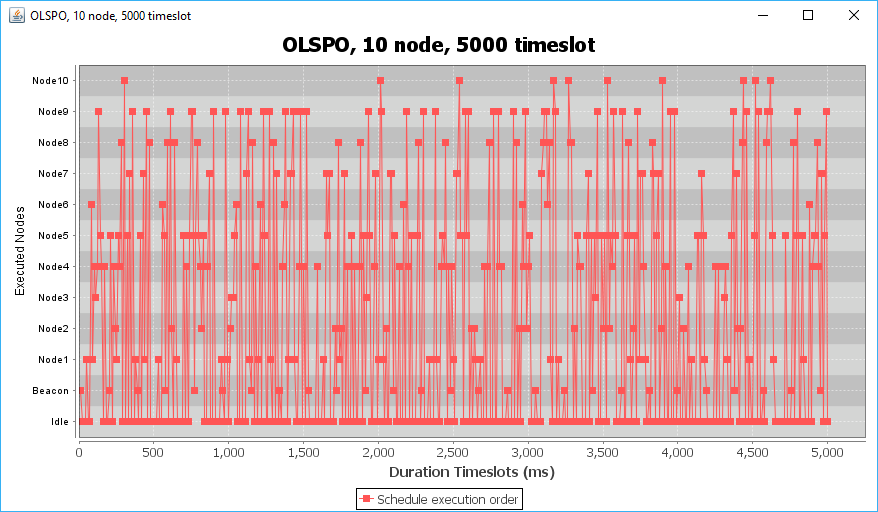} \\
	\caption{Simulation: OLPSO with 500 timeslot and 10 nodes}
	\label{fig:12}
\end{figure}

\begin{figure}
	\centering
	\includegraphics [width=0.48\textwidth]{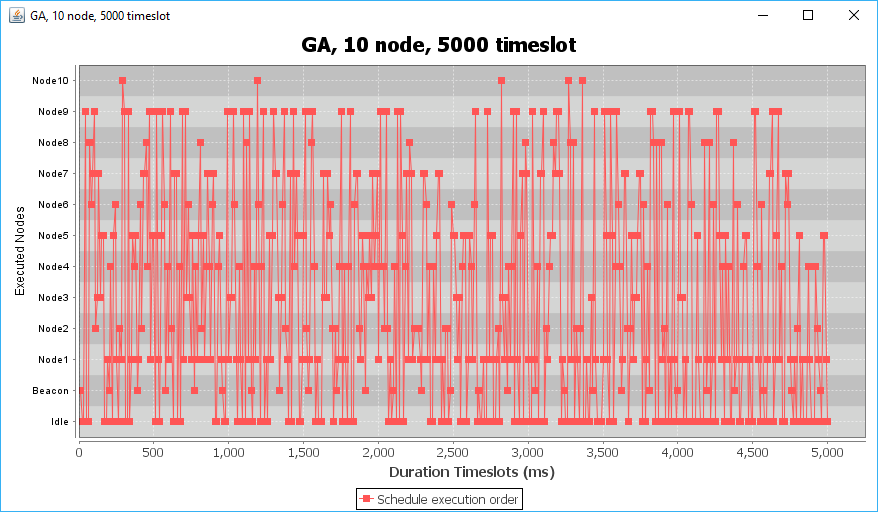} \\
	\caption{Simulation: GA with 500 timeslot and 10 nodes}
	\label{fig:13}
\end{figure}

\subsection{Simulation result for 100 nodes}
\begin{figure}
	\centering
	\includegraphics [width=0.48\textwidth]{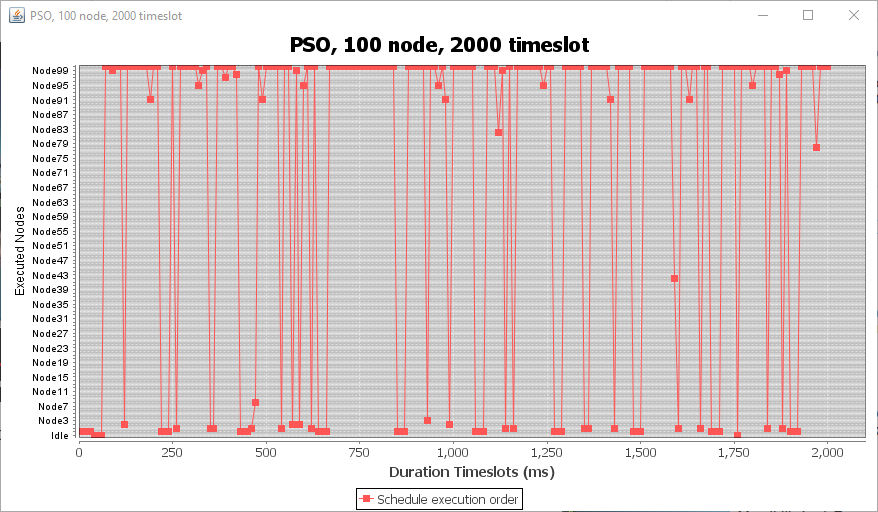} \\
		\caption{Simulation: PSO with 200 timeslot and 100 nodes}
	\label{fig:15}
\end{figure}

\begin{figure}
	\centering
	\includegraphics [width=0.48\textwidth]{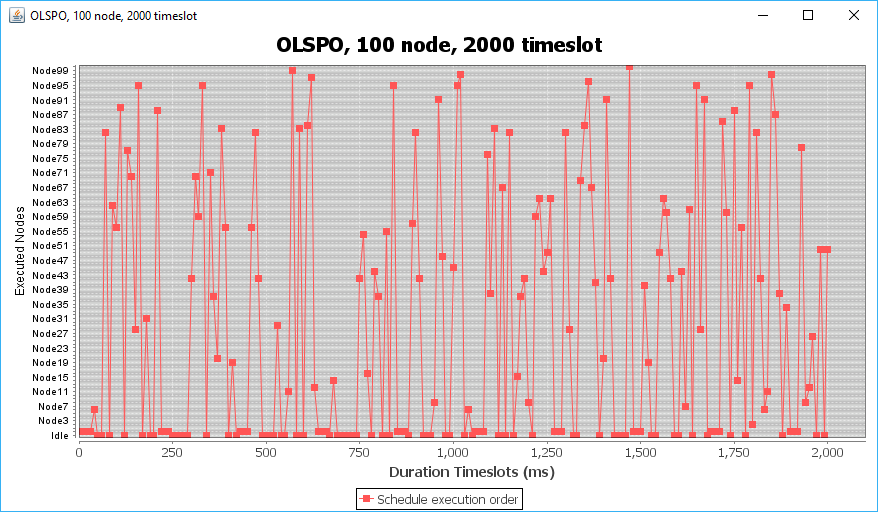} \\
	\caption{Simulation: OLPSO with 200 timeslot and 100 nodes}
	\label{fig:16}
\end{figure}

\begin{figure}
	\centering
	\includegraphics [width=0.48\textwidth]{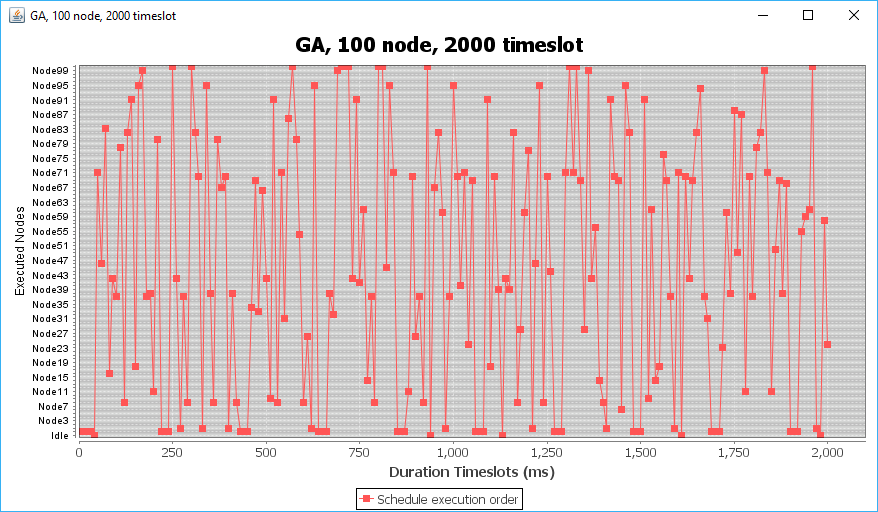} \\
	\caption{Simulation: GA with 200 timeslot and 100 nodes}
	\label{fig:17}
\end{figure}

\begin{figure}
	\centering
	\includegraphics [width=0.48\textwidth]{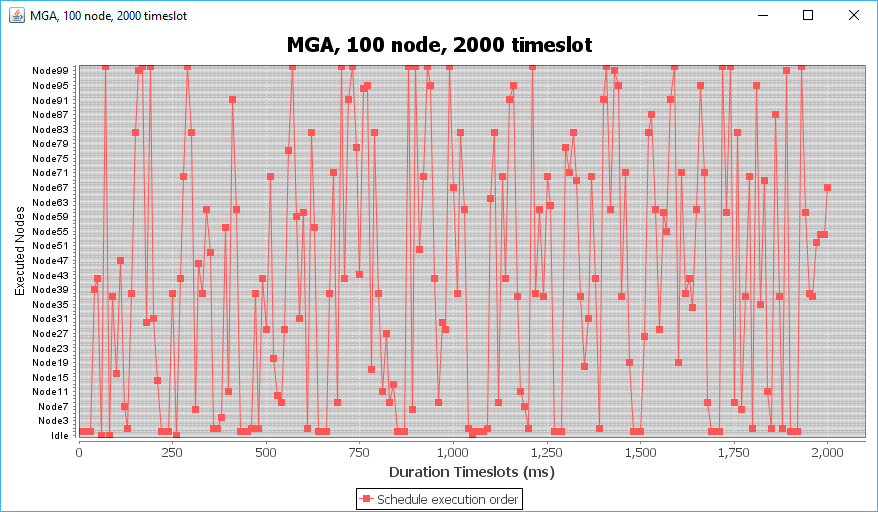} \\
\caption{Simulation: MGA with 200 timeslot and 100 nodes}
	\label{fig:18}
\end{figure}

Figs. \ref{fig:15}, \ref{fig:16}, \ref{fig:17}, and \ref{fig:18} show a case study with a sufficiently large number of sensors that is 100 SN. We test this 100 SN in 3 categories 100 ts, 200 ts, and 500 ts. This sub-section would describe the 200 ts for 100 SN. Fig. \ref{fig:15} was the result for PSO, Fig. \ref{fig:16} was the result for OLPSO, Fig. \ref{fig:17} was the result for GA, while Fig. \ref{fig:18} was the result for MGA.

Metaheuristic algorithms are able to find the optimum solution in a search process in real time to generate better solutions \cite{29}. We used PSO and GA in this paper. In addition, we made improvements for GA that we called the MGA. PSO in this scenario is still less able to provide an optimum solution when working with timeslots. The idle time value for PSO of 11 ts (110ms) is far more than GA 8 ts (80 ms) and MGA 3 ts (30ms). Indeed our fitness calculations are not only based on idle time, but also the number of miss-deadlines. PSO is still less than optimal for case 100 SN.

For the large scale experiment in this scenario which used 100 SN, the OLPSO still performed with almost the same performance as in the 10 SN scenario. It also enhanced the evidences that OLPSO is less suited for superframe scheduling case on ISA 100.11a. This is due to the characteristics of OLPSO which are vector based as in traditional PSO while this case is TDMA-superframe based (discrete based). Although OEM was proposed to solve problems on local optima, in this case OLPSO still produced a premature coverage that showed the existence of some idle timeslots.

From Fig. \ref{fig:17} and \ref{fig:18}, GA and MGA are more adaptive to this combinatorial and discrete chassis. In accordance with the explanations in section I, PSO and GA are the same metaheuristic algorithms, but have different search space solutions. For GA in each iteration there will always be an optimum solution, because of the three components (selection, mutation, and crossover). The PSO algorithm uses particles and velocities which follow the direction of the fitness value. However, for case optimization of timeslots on this IWSN, PSO is often hampered in discovering the global optimum solution, and this is seen from the idle-time results, as previously discussed.

\section{Results and summary}
\label{5}

\begin{table*}
	\centering
	\caption{Scenarios \& simulation results}
	\label{tab:13}
	\begin{tabular}{ | c | c | c | c | c | c | c | c | c | } \hline
		\multirow{2}{*} {\textbf{No}} & \multirow{2}{*} {\textbf{No. of Ts}}   & \multirow{2}{*} {\textbf{No. of Node}}  & \multicolumn{6}{ c|}{Defect time (ms)} \\ \cline{4-9}
		& & & \textbf{EDF \cite{30}} & \textbf{DMS \cite{6}}  & \textbf{PSO} & \textbf{OLPSO \cite{43}} & \textbf{GA}  & \textbf{MGA}\\ \hline \hline
1      & 100    & 4    & 2220   & 2180    & 2214  	& 2199   	& 2100  	& 2010   \\ \hline 
2      & 100    & 7    & 3090   & 3330    & 3321  	& 3457  	& 2764  	& 2279   \\ \hline 
3      & 100    & 10   & 6770   & 5080    & 6391  	& 6940  	& 5071  	& 4898   \\ \hline 
4      & 100    & 100  & 84970  & 85070   & 82961 	& 83295  	& 79688 	& 79703  \\ \hline 
5      & 200    & 4    & 4720   & 4520    & 4834  	& 5237  	& 4366  	& 4062   \\ \hline 
6      & 200    & 7    & 7190   & 7120    & 8025  	& 8388  	& 6054  	& 4882   \\ \hline 
7      & 200    & 10   & 14850  & 11590   & 16071 	& 16728 	& 12463 	& 10901  \\ \hline 
8      & 200    & 100  & 184050 & 184220  & 180745	& 182236	& 174675	& 174427 \\ \hline 
9      & 500    & 4    & 11800  & 11360   & 13673 	& 14555 	& 11876 	& 10432  \\ \hline 
10     & 500    & 7    & 17190  & 18300   & 22977  	& 24002 	& 17355 	& 14015  \\ \hline 
11     & 500    & 10   & 39480  & 29020   & 44182  	& 46203 	& 36266 	& 28263  \\ \hline 
12     & 500    & 100  & 481130 & 481640  & 475533	& 480237	& 471593	& 465529 \\ \hline 
	\end{tabular}
\end{table*}

\begin{figure}
	\centering
	\includegraphics [width=0.48\textwidth]{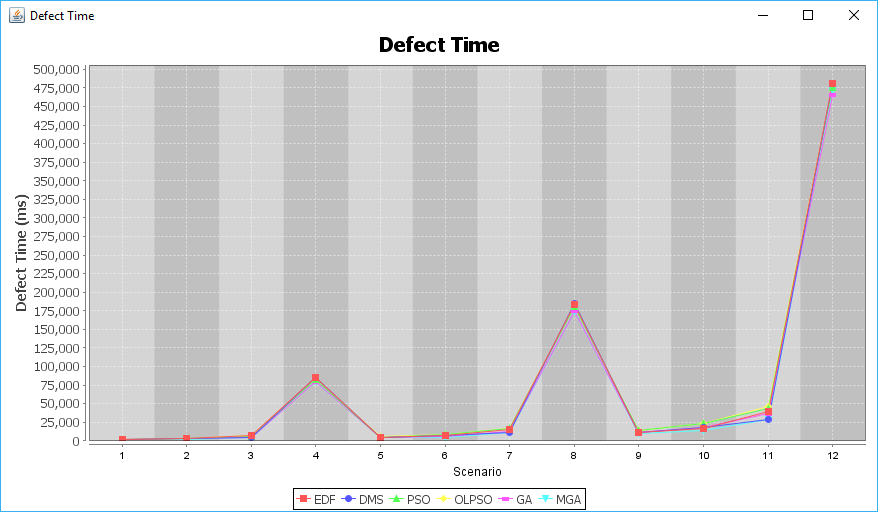} \\
	\caption{Simulation results summary: defect time}
	\label{fig:19}
\end{figure}

\subsection{Analysis of defect time}

The PSO and OLPSO modeled here consists of a swarm of 50 particles initialized with a population of random candidate solutions. Each particle has a position represented by position vector and a velocity represented by velocity vector. For fitness values, we use $c_1$ as a positive constant, called coefficient of self-recognition component, and $c_2$ as the positive constant, called coefficient of social component. The results from 60 scenarios produced by PSO and OLPSO are better than by DMS, but worse than by GA, because the searching process of \textit{global minima} on PSO and OLPSO is continuous, while the case of timeslot scheduling on IWSN is a discrete case.

GA is an algorithm which seeks the best solution from the population based on natural gene selection and recombination. The recombination is done in a random process. The result of the defect-time calculation is obtained from crossover and mutation from a few of the best candidates from the population. It really depends on the crossover probability ($p_c$) and the mutation rate ($p_m$) value. In short, the defect-time values heavily depend on the attributes of GA and MGA that will be set or used. Based on our experiments, the optimum configurations for IWSN case study are as shown in Table \ref{tab:3}.

\subsection{Analysis of complexity and stability}

\begin{figure}
	\centering
	\includegraphics [width=0.48\textwidth]{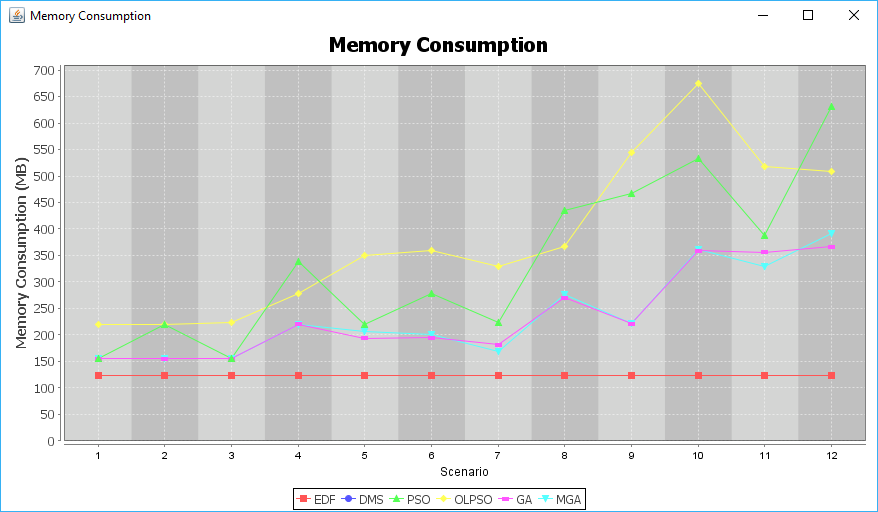} \\
	\caption{Simulation results summary: memory consumption}
	\label{fig:20}
\end{figure}

\begin{figure}
	\centering
	\includegraphics [width=0.48\textwidth]{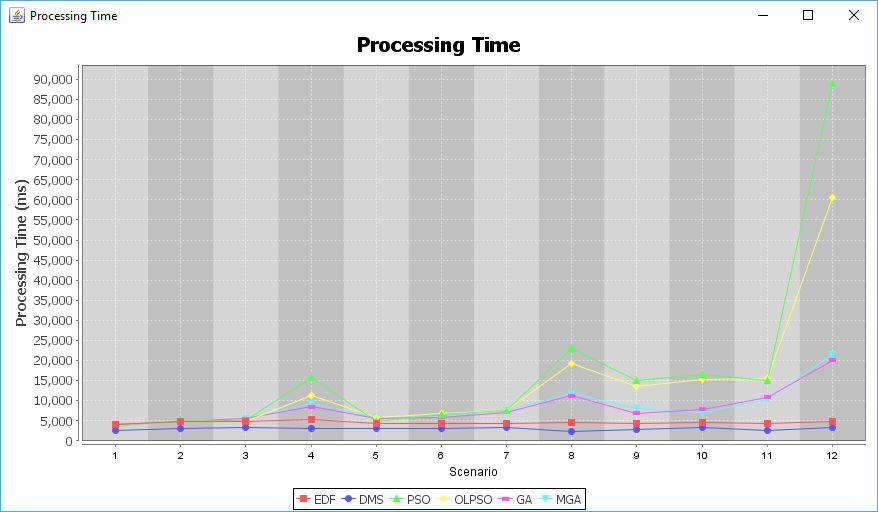} \\
	\caption{Simulation results summary: processing time}
	\label{fig:21}
\end{figure}

Figs. \ref{fig:20} and \ref{fig:21}, illustrate complexity and stability comparison in terms of memory consumption and processing time respectively. It can be seen that EDF and DMS comparatively demonstrate lower complexity and higher stability, considering its small memory consumption and less processing time. This is because EDF and DMS are non-metaheuristic algorithms. The impact of computational complexity and time stability on the application of evolutionary computing approach in superframe scheduling at IWSN i.e., PSO, OLPSO, GA, and MGA, is explained para wise as follows.

As shown in Fig. \ref{fig:20}, which presents the case study of 60 scenarios, GA and MGA demonstrate less complexity compared to PSO and OLPSO. To the best of authors knowledge, there are three reasons behind the lower complexity of GA and MGA. Firstly, PSO and GA have different calculations for producing new population; calculation of PSO is based on speed and direction, whereas GA uses crossover and mutation. Secondly, the process of generating new solutions (offspring) in PSO and GA is based on fitness value, as described in the Sec. \ref{3}. Lastly, through more insight, it can be seen that PSO always compare its current best solution with the previous one and keeps the best of the two, whereas in GA, there is no process of storing. The GA directly performs crossover and mutation process of the gene that has low fitness.

The stability analysis is used for calculating the processing time of each scenario. As can be seen in Figure \ref{fig:21}, GA and MGA have faster processing time than PSO and OLPSO. Some primary reasons causing GA and MGA to have good processing time are detailed as follows. Firstly, because the PSO is vector-based optimization, if one or more of the neighbors assume that the \textit{i}-th solution (even premature) is best, this is often referred to as trapped within local optima. Secondly, this case study proves that PSO always produce homogeneous solutions (shown in Fig. \ref{fig:15}); several generations of defect time referred to the same values in several timeslots. Lastly, the equations of PSO and OLPSO have more detailed parameters in the searching process, but the authors conclude that this feature is not suitable to be implemented on discrete cases. In summary, GA and MGA have better processing time than PSO and OLPSO for scheduling superframe in IWSN.

\subsection{Validation of the algorithms}
There are 12 different types of data scenarios seen in Table \ref{tab:13} with each tested with PSO, OLPSO, GA, and MGA respectively. In \cite{29} chapter 9 in particular, it is said that all algorithms belonging to metaheuristics will produce different solutions for each algorithm being executed. Therefore, to ensure validation of all such valid heuristic algorithms the iteration must be repeated between 10 and 100 times. In this paper, repetition of 10 times is used as recommended in \cite{29}. It aims to ensure a high success rate in finding a global minima and that a report of fitness value is not met by chance. For the validation of these four algorithms, several test cases are enclosed in Appendix \ref{app:1}. For each test we also calculate the mean and standard deviation ($\sigma$) as the measurements.

We do not repeat iterations for DMS and EDF because they are belong to non-metaheuristic methods, which in its calculations use the priority assignment policy \cite{30}\cite{31}. If a repetitive iteration was done on DMS and EDF, it will produce the same value, and vice versa on metaheuristic algorithms.

\subsection{Brief summary of results}
Table \ref{tab:13} shows the results of all the experiments that have been carried out. With variations in timeslots, wireless sensor nodes, and algorithms that we used, there are 60 experiments in total. Each experiment takes approximately 30 minutes per simulation. The authors propose using GA and MGA for optimizing superframe scheduling on IWSN (especially ISA 100.11a). These proposals are strengthened by comparing EDF \cite{30} \& DMS \cite{6} as representatives of non-metaheuristic methods with PSO \& OLPSO as representatives of metaheuristic methods. The simulation results show that MGA is better than conventional EDF, DMS, PSO, OLPSO and GA. GA is a metaheuristics algorithm with two components: exploitation for searching and exploration for generating the solutions. Therefore, we need to be careful in setting the parameters and the fitness value.

\begin{table}[]
\centering
\scriptsize
\caption{summary of performance evaluation}
\label{tab:14}
\begin{tabular}{ | p{1.3cm} | p{.7cm} |  p{.7cm} | l | l | l |  p{.7cm} |} \hline
\textbf{Metrics}  	& \textbf{EDF} & \textbf{DMS} 	& \textbf{PSO} 	& \textbf{OLPSO} 	& \textbf{GA} 	& \textbf{MGA}	\\ \hline \hline
Defect time 		& Fair 		   & Fair 			& Good			& Fair				& Good			& Very Good		\\ \hline
Memory consumption  & Very Good	   & Very Good		& Poor			& Poor				& Good			& Good			\\ \hline
Processing Time  	& Very Good	   & Very Good		& Poor			& Poor				& Good			& Good			\\ \hline
\end{tabular}
\end{table}

Based on the three results of analyses and validation of proposed algorithm, we provide conclusions for the experiments of the performance of each algorithm in Table \ref{tab:14}. The credit score that we provide are categorized as bad, poor, fair, good, or very good. The evaluation was taken based on the average of 12 experiments that have been conducted. To the best of author's knowledge, GA or MGA can be applied in IWSN especially for smart factory.

\section{Future work and extensions}
\label{6}

The work presented in this paper is the initial work that has considered GA for scheduling in the industrial networks. Therefore, we have considered a basic scenario for all findings. Since this work is based on heuristic approach, new parameters and environments are required to prove the supremacy of this algorithm in industrial networks with diverse scenarios. Thus, authors have planned to work further on this approach with future directions as mentioned below.

Based on system scenarios that have been established using offline testing, it will be valuable if GA can be applied to the real system. For large-scale testing, it is possible to try it using thousands of sensor nodes and to use more than one gateway. There are still needs of improvisation for defect time, complexity, and stability based on Table \ref{tab:13}. The configuration in this paper is specifically for scheduling on the IWSN problems. For other optimization problem, particular configuration of attributes is required to obtain a better solution. Furthermore, it can be used in combination with other optimization algorithms.

\section{Conclusion}
\label{7}

This paper successfully demonstrated the use of the presented methods for DMS problems in scheduling superframes. As an alternative solution and enhancement to current methods, we recommend the use of GA for optimizing scheduling in IWSN superframes. We have presented GA and MGA by using Java programming for ease of development, and because it is supported by multiple platforms. We have presented PSO and OLPSO as a comparative study with EA. From 60 scenarios tested above, the PSO and OLPSO are not suitable for discrete cases, however PSO and OLPSO are better for continuous cases. Timeslot optimization in IWSN, as used in this paper, is an example of a discrete case. This paper also provides recommendations for the configurations of GA attributes that need to be set. This paper concludes that MGA is the best in optimizing the defect time for scheduling superframes in IWSN, especially in ISA 100.11a. This supports the theory that GA can provide a solution for superframe scheduling problems in IWSN.

\section*{Acknowledgment}
Acknowledgment.

\appendices

\section{Testing accuracies PSO, OLPSO, GA, and MGA}
\label{app:1}

The 12 scenarios that have been explained in Section \ref{5} will also be run on these four algorithms. Below are the results of four out of those 12 scenarios, which are scenario 3, 6, 9 and 12.

\begin{table}[h]
	\centering
	\scriptsize
	\caption{Scenario 3}
	\label{tab:23}
	\begin{tabular}{ | c | c | c | c | c | c |}
		\hline
	Test Case & Experiment &  \multicolumn{4}{c|}{Defect Time (ms)}   \\ \hline
		&  		& PSO 	& OLPSO & GA & MGA  \\ \hline 
3 	& 1 	& 6600 & 6660 & 5520 & 4830 \\ \hline                         
	& 2 	& 6160 & 6940 & 5010 & 4910 \\ \hline                         
	& 3 	& 6880 & 6660 & 4690 & 4890 \\ \hline                         
	& 4 	& 6340 & 7000 & 4890 & 4910 \\ \hline                         
	& 5 	& 6090 & 7070 & 4910 & 4840 \\ \hline                         
	& 6 	& 6410 & 7460 & 5340 & 4910 \\ \hline                         
	& 7 	& 6250 & 6890 & 5210 & 4980 \\ \hline                         
	& 8 	& 6290 & 6540 & 4980 & 5080 \\ \hline                         
	& 9 	& 6640 & 6960 & 5090 & 4810 \\ \hline                         
	& 10 	& 6250 & 7220 & 5070 & 4820 \\ \hline                         
Mean &  	& 6391 & 6940 & 5071 & 4898 \\ \hline                         
$\sigma$ &  & 245.33197 & 276.68674 & 237.92856 & 83.10635	 \\ \hline  
	\end{tabular}
\end{table}

\begin{table}[h]
	\centering
	\scriptsize
	\caption{Scenario 6}
	\label{tab:26}
	\begin{tabular}{ | c | c | c | c | c | c |}
		\hline
		Test Case & Experiment &  \multicolumn{4}{c|}{Defect Time (ms)}   \\ \hline
		&  		& PSO 	& OLPSO & GA & MGA  \\ \hline
6 		& 1  & 8060 & 8740	& 6240 & 4960  \\ \hline                       
		& 2  & 8070 & 8270	& 6180 & 4930  \\ \hline                       
		& 3  & 8010 & 8460	& 6120 & 4810  \\ \hline                       
		& 4  & 8310 & 8090	& 6070 & 4990  \\ \hline                       
		& 5  & 8570 & 8750	& 5920 & 4820  \\ \hline                       
		& 6  & 7460 & 8130	& 5970 & 4850  \\ \hline                       
		& 7  & 8060 & 8500	& 6140 & 4910  \\ \hline                       
		& 8  & 8150 & 8130	& 5850 & 4810  \\ \hline                       
		& 9  & 7630 & 8300	& 6040 & 4870  \\ \hline                       
		& 10 & 7930 & 8510  & 6010 & 4870  \\ \hline                       
Mean 	&    & 8025 & 8388	& 6054 & 4882  \\ \hline                       
$\sigma$ &  & 313.20032 & 242.93574 & 120.75687 & 63.56099	 \\ \hline 
	\end{tabular}
\end{table}

\begin{table}[h]
	\centering
	\scriptsize
	\caption{Scenario 9}
	\label{tab:29}
	\begin{tabular}{ | c | c | c | c | c | c|}
		\hline
		Test Case & Experiment &  \multicolumn{4}{c|}{Defect Time (ms)}   \\ \hline
		&  		& PSO 	& OLPSO & GA & MGA  \\ \hline
9 		& 1  & 14000 & 14550	& 12000 & 10520  \\ \hline                   
		& 2  & 13500 & 14470	& 11700 & 10540  \\ \hline                   
		& 3  & 13700 & 14270	& 11990 & 10310  \\ \hline                   
		& 4  & 13820 & 15050	& 12090 & 10390  \\ \hline                   
		& 5  & 13620 & 14520	& 12130 & 10330  \\ \hline                   
		& 6  & 14040 & 14720	& 11840 & 10400  \\ \hline                   
		& 7  & 13130 & 14110	& 12020 & 10420  \\ \hline                   
		& 8  & 13990 & 14900	& 11680 & 10490  \\ \hline                   
		& 9  & 13320 & 14350	& 11650 & 10510  \\ \hline                   
		& 10 & 13610 & 14610 	& 11660 & 10410  \\ \hline                    
Mean 	&  	 & 13673 & 14555	& 11876 & 10432  \\ \hline                   
$\sigma$ &  & 301.95842 & 283.63709 & 190.85771 & 79.97221	 \\ \hline 
	\end{tabular}
\end{table}

\begin{table}[h]
	\centering
	\scriptsize
	\caption{Scenario 12}
	\label{tab:32}
	\begin{tabular}{ | c | c | c | c | c | c |}
		\hline
		Test Case & Experiment &  \multicolumn{4}{c|}{Defect Time (ms)}   \\ \hline
		&	 & PSO 	  & OLPSO   & GA     & MGA  \\ \hline    
12		& 1  & 474990 & 480370	& 464330 & 465450   \\ \hline                     
		& 2  & 475930 & 480160	& 472360 & 466490   \\ \hline                     
		& 3  & 474730 & 480590	& 471850 & 465080   \\ \hline                     
		& 4  & 475750 & 481480	& 472240 & 465140   \\ \hline                     
		& 5  & 476000 & 479630	& 473200 & 465290   \\ \hline                     
		& 6  & 476040 & 480020	& 472880 & 465340   \\ \hline                     
		& 7  & 475470 & 479790	& 472210 & 466460   \\ \hline                     
		& 8  & 474220 & 480230	& 471870 & 465030   \\ \hline                     
		& 9  & 476830 & 480350	& 472110 & 465130   \\ \hline                     
		& 10 & 475370 & 479750  & 472880 & 465880   \\ \hline                     
Mean &   & 475533 & 480237	& 471593 & 465529   \\ \hline                     
$\sigma$ &  & 749.53392 & 533.18852 & 2591.52486 & 555.14662	 \\ \hline
	\end{tabular}
\end{table}



\bibliographystyle{unsrt}
\bibliography{reference}

%




\end{document}